# Gate Tunable Asymmetric Ozone Adsorption on Graphene

Zhen Qi[1,2,*], Wanlei Li[1,2,*], Jun Cheng[1,2,*], Zhongxin Guo[1,2], Chenglong Li[1,2], Shang Wang[1,2], Zuoquan Tan[1,2], Zhiting Gao[3], Yongchao Wang[3], Zichen Lian[3], Shanshan Chen[1,2], Yonglin He[4], Zhiyong Wang[4], Yapei Wang[4], Jinsong Zhang[3,5,6], Yayu Wang[3,5,6], Peng Cai[1,2,6,†]

[1]*Department of Physics and Beijing Key Laboratory of Opto-electronic Functional Materials and Micro-nano Devices, Renmin University of China, Beijing 100872, China*

[2]*Key Laboratory of Quantum State Construction and Manipulation (Ministry of Education), Renmin University of China, Beijing, 100872, China*

[3]*State Key Laboratory of Low-Dimensional Quantum Physics, Department of Physics, Tsinghua University, Beijing 100084, China*

[4]*Key Laboratory of Advanced Light Conversion Materials and Biophotonics, School of Chemistry and Life Resources, Renmin University of China, Beijing 100872, China*

[5]*Frontier Science Center for Quantum Information, Beijing 100084, China*

[6]*Hefei National Laboratory, Hefei 230088, China*

[*]*These authors contributed equally to this work.*

[†]pcai@ruc.edu.cn

**Molecular adsorption is pivotal in device fabrication and material synthesis for quantum technology. However, elucidating the behavior of physisorption poses technical challenges. Here graphene with ultrahigh sensitivity was utilized to detect ozone adsorption at cryogenic temperatures. Significant hole doping observed in graphene indicates a strong interaction between ozone and graphene. Interestingly, the adsorption exhibits asymmetry with positive and negative gate voltages. The strong affinity of ozone provides a tool to modulate materials and devices, while the gate tunability of adsorption offers new insights into construction and manipulation of oxide quantum materials.**

Adsorption processes play an essential role across diverse scientific and technological fields, including advanced material synthesis, chemical/biomedical engineering, environmental remediation, energy storage/conversion and quantum technology. A fundamental understanding and manipulation of adsorption behaviors on surface is crucial for developing novel functional materials[1–7] and next generation of quantum devices[8,9]. In presence of various interactions, the kinetics and efficiency of chemical and physical adsorption are closely related to complex parameters, such as surface/interface[10], absorbates, temperature[11], and external fields[12]. However, a complete understanding of the mechanism of these tuning parameters remains elusive, especially for the electric-related adsorptions in electrochemical processes. Moreover, precise control and investigation of adsorption are still extremely challenging, since the detection of surface /interface is technically difficult, especially at low temperatures[13].

An ideal sorption example is the ozone adsorption on graphene in a stable carbon honeycomb lattice structure[14]. On the one hand, with its two-dimensional atomic layer thickness, the bulk properties of graphene can directly reflect its surface mediated by adsorption. In the presence of Dirac electrons, charge neutral point (CNP) of graphene captured traditional transport experiments exhibits ultrahigh sensitivity to molecular adsorption, resulting in charge transfer between absorbates and graphene[15]. On the other hand, The adsorption of polar molecules ozone is widely used in surface reaction[16–18] and controlling oxide functional materials[19,20]. By ozone adsorption, the oxygen content of two-dimensional cuprate superconductor can be adjusted, resulting in regulating effective doping concentration and fine-tuning of exotic high-temperature superconductivity[21]. Additionally, ozone adsorption has been found to functionalize graphene in assistance of ultraviolet light or elevated temperatures, which can alter its electrical and optical properties, thereby expanding the potential applications.

The ozone adsorption process on graphene could be classified into chemisorption and physisorption. In chemisorption, the formation of an epoxy group, wherein an oxygen atom bonds to two neighboring carbon atoms without breaking the carbon-carbon bond, is the most predominant and intriguing chemical bonding[14,17,22]. Before forming chemical bonds, ozone molecules are believed to exist in stable physisorption state, wherein ozone are physically attracted to the surface without forming chemical bonds, from a computational perspective[14,23,24]. As depicted in Fig. 1a, an energy barrier exists between physisorption and chemisorption, because of increased repulsive energy as ozone get closed to graphene. As revealed in Fig. 1b, these possible stable physisorption states can surpass the energy barrier and transit into chemisorption upon elevated temperature and ultraviolet light irradiation[25–28]. However, the study of physisorption is scarce from both theoretical and experimental perspective. Since physisorption is believed to dominate at low temperatures, here we use graphene as sensor to study the ozone sorption at low temperatures.

Using its CNP as a detector, the sensitivity of graphene is governed by the homogeneity of its lattice and electronic structure. To achieve a more homogeneous graphene layer, the flatness of graphene can be enhanced by placing it on a hexagonal boron nitride (h-BN) crystalline substrate. Additionally, the graphene homogeneity can be further preserved by using metallic gating electrodes with thick layers of graphite. To accurately measure the evolution of the CNP as ozone is absorbed on graphene layer, the top graphene surface should be exposed and a standard four-probes method resistance measurement is employed for graphene layer using hBN as gate dielectric layer, as schematically shown in Fig. 1c. A thick layer of graphite can also serve as an auxiliary electrode to reduce the contact resistance between the h-BN and the electrode. The stacking of the graphene/h-BN/graphite multilayer can be achieved using traditional transfer technology[29], to ensure the top graphene layer is exposed. A typical stacked device with hBN layer ($d_B$ = 17nm) on $SiO_2$ ($d_{SiO2}$ = 285 nm) is depicted in Fig. 1d, before electric contacts are fabricated. Either before or after measurements, the graphene layer can be confirmed by checking the G/2D peak in Raman spectroscopy, as revealed in Fig. 1e.

At cryogenic temperature T = 200 K, the CNP position of graphene shows a significant response as ozone gas is introduced, as revealed in Fig. 2a. For better comparison, the CNP of pristine graphene layer at 200 K is determined in vacuum before exposure to ozone, by measuring the evolution of electric resistance upon gating voltage (R-$V_G$ curve)[30]. The resistance is highest as the Fermi level of graphene is tuned to CNP by gating voltage. After the measurement, ozone mixture was introduced into sample chamber at 200 K (well above ozone boiling temperature 161 K) for five minutes, during which the pressure was kept at 10 mbar and the gating voltage stays at 0 V. After ozone in the chamber is purged, the R-$V_G$ of graphene layer is carefully measured. The shift of CNP reveals hole doping the graphene, indicated the ozone mixture are absorbed on graphene layer. Moreover, the adsorption can be roughly controlled by exposure time as shown in Fig. 2a. With reference to the CNP of the pristine sample, the shift of CNP increases with prolonged ozone exposure with same pressure. Hereafter, without further specified, the ozone exposure time is 5 minutes during which the pressure was kept at 10 mbar and the gating voltage stays at 0 V.

The CNP shift of the graphene layer is attributed to adsorption of ozone rather than oxygen. While oxygen is predominant in the gas mixture, it can also be dissociated from ozone. However, the possibility of CNP shift due to oxygen can be ruled out, as no significant CNP shift is observed in Fig. 2b when the graphene layer is exposed to pure oxygen gas under the same conditions, even for 20 minutes. This result demonstrates oxygen hardly reacts with graphene at 200 K, resulting from extremely low chemical reaction rate at cryogenic temperature. The CNP shift remained stable throughout the experiment, suggesting that ozone dissociation is sluggish at 200 K, and the physically absorbed ozone can exist in molecular form during our experiments.

The remaining issue is to analyze the effect of ozone adsorption on the CNP shift $\Delta V_{CNP}$, since both chemisorption and physisorption of ozone are considered to modulate the carrier density of graphene layer[17,25,26,26,28,31], yet remain inconclusive. Figure 2c shows the change of carrier density $\Delta n$ induced by ozone adsorption, by $n = \frac{\varepsilon\varepsilon_0 V_G}{te}$, where $\varepsilon_0$ is dielectric constant and t is the thickness of hBN layer[30]. Before analysis, it is worth noting that detecting physisorption is challenging due to its inherent experimental difficulty, whereas chemical adsorption could be detectable because it involves the formation of chemical bonds, such as C-O bonds. Ultraviolet (UV) irradiation decomposes ozone molecule into oxygen molecule and active oxygen atom, in which active oxygen atom reacted with carbon atoms on the surface of graphene, thus introducing hole carriers doping on graphene layer due to oxygen can suck two electrons from graphene layer at most. However, this chemisorption rate at room temperature without UV irradiation is extremely limited, which cannot lead to such large Δn in graphene. In the literature[23], the predominant C-O bond coverage carefully detected by high-resolution photoemission spectroscopy, while the graphene is exposed to ozone with 1 bar pressure for several hours. Fig. 2c also shows the estimated Δn derived from this C-O coverage on graphene if we simply assume each O atom can suck 2 electrons (1 electron is trapped in each C-O bond) for overestimation. Noting that, ozone pressure is 1 bar while the ozone pressure is only 10 mbar. Even though the carriers change due to C-O bond is far overestimated, it is just in the same order of magnitude of Δn due to ozone adsorption at 200 K, suggesting that only chemisorption cannot lead to the large Δn at 200 K and physisorption play a significant role in modulating carrier density of graphene.

Owing to extremely low reaction rate and technically challenges at cryogenic temperature, the coverage of C-O bond in chemisorption at 200 K cannot be experimentally determined, however, it could be approximately estimated by extrapolating from the adsorption coverage at elevated temperatures, based on Eyring equation[23] in adsorption theory. The temperature dependence of C-O bond coverage $\theta$ formed in the ozone chemisorption approximately follows the derived formula[23], $\theta \propto k_c = \frac{k_B T}{h} \exp\left(\frac{-E_c}{k_B T}\right)$, where $E_c$ refers to the energy barrier for chemisorption, $k_B$ is Boltzmann constant, $h$ is Planck's constant. As shown in Fig. 2c, the carrier density derived from the extrapolated C-O bonds coverage at 200 K is several orders of magnitude smaller than the observed Δn. The conclusion is similar regardless of whether the extrapolation parameter is obtained by linear or exponential fitting[23]. At cryogenic temperatures, other types of bonds, such as C=O bonds, in chemisorption can be further disregarded due to their higher energy barrier, resulting in a significantly lower reaction rate. Again, it is important to note that the C-O bond coverage around room temperature is obtained upon ozone exposure (1 bar). It indicates that the observed Δn in graphene upon ozone exposure (10 mbar) mainly originates from physisorption rather than chemisorption.

To check whether dielectric materials influence the ozone adsorption, similar experiments were carried on the graphene layer directly sitting on $SiO_2$ dielectric layers, to replace hBN with $SiO_2$ (285 nm thickness). Under similar ozone exposure conditions, the change in carrier density Δn of the graphene layer remains nearly the same, despite a significantly higher gating voltage required to reach the CNP. The ozone adsorption on graphene is similar in these two cases. Additionally, the Δn increases as adsorption temperature is elevated while the other parameters stay the same, indicating the adsorption is dependent of temperature.

Figure 2d depicts the schematic electronic structure modulation of graphene as ozone is physically absorbed. On one hand, the decrease of Dirac electron density in graphene could possibly ascribed to charge transfer between graphene and ozone molecules, similar to Van der Waals type heterostructures[32,33] . On the other hand, the lowering of Fermi level in graphene device might result from the collective polarity of ozone molecules in specific configuration on graphene layer. These two possibilities cannot be distinguished here.

To gain deeper insight into the relationship between ozone adsorption and the electronic state of graphene, experiments were conducted on graphene layers with varying initial CNPs. Figures 2e and 2f illustrate the CNP shifts upon adsorption for pristine graphene layers with electron and hole doping, respectively. Different types of intrinsic doping were deliberately characterized for the graphene samples prior to ozone exposure at 200 K. In both cases, ozone adsorption consistently resulted in the introduction of hole-type carriers into the graphene layers. The results revealed similar levels of hole doping, albeit with slight variations among the different samples.

Alternatively, gate voltage can be applied to tune the carrier density of graphene layer, offering an operando stage to study the relationship between ozone adsorption and the electronic state of graphene. The adsorption is conducted with constant bottom gate voltage applied to dielectric hBN at 200 K, while the graphene layer is grounded. Immediately after completing the ozone exposure, the applied gate voltage is removed, and the R-Vg curve of the graphene layer, as shown in Fig. 3a, is measured to determine the CNP position. Figure 3b depicts the evolution of $V_{CNP}$ before and after ozone exposure in the sequence of applying gate voltage. The change in carrier density Δn, as shown in Fig. 3b, is derived from the change in $V_{CNP}$ before and after the adsorption. Figure 3c summarizes the gate dependence of Δn, indicating that the adsorption is strongly influenced by the gate voltage. A negative gate voltage may exert an inhibitory effect on this process. This result explicitly reveals that the effect of positive and negative gate voltage on adsorption is asymmetric at the hole doping level. In this measurement, the same graphene layer is continuously modulated and the ozone adsorption is accumulated.

To mitigate the accumulation effect, attempts were made to remove ozone after each adsorption cycle. The graphene layer was restored close to its initial state through vacuum annealing at room temperature. The experiment was conducted on graphene layer atop dielectric $SiO_2$ layer. This choice was made to facilitate better monitoring of the graphene's restoring to its initial state, given the thicker $SiO_2$ layer requiring a larger gate voltage. Figures 4a and 4c display the R-$V_G$ curves before and after ozone exposure, with a gate voltage of +10 V and -10 V, respectively. Remarkably, the gate-tunable asymmetric adsorption remains consistent. The Δn is enhanced after ozone exposure with the application of positive voltage, indicating stronger ozone adsorption. Conversely, the Δn is suppressed after ozone exposure with the application of negative voltage, indicating weaker ozone adsorption. As depicted in Fig. 4e and 4f, this phenomenon was further validated by repeating the adsorption cycles with the application of gating voltage. Moreover, it was reproduced across different samples. The detailed evolution varied in different cycles or different samples, possibly due to some irreversible complex processes and intrinsic inhomogeneity in each sample. However, the primary conclusion remains unchanged: the adsorption is asymmetrically dependent on the gating voltage, regardless of the order of voltage applied here. The intriguing universality of ozone adsorption on graphene persists across various conditions and substrates.

Figure 4b and 4d illustrate the schematic process as the gate voltage is applied. When a positive voltage is applied, more electrons are injected into the graphene layer due to electrostatic doping, while electrons are extracted from the graphene when a negative voltage is applied. This suggests that the electronic state plays a crucial role in the adsorption process. Interestingly, there is no significant difference in ozone adsorption between graphene layers with intrinsic electron doping and those with intrinsic hole doping. Thus, the additional carriers tuned by the gate voltage may be essential for the asymmetric adsorption behavior.

Whether the hole doping introduced by ozone arises from charge transfer or the collective polarity of ozone molecules, the ozone configurations on graphene significantly influence the interaction between Van-der-Waals molecules and graphene. Previous calculations have shown that different ozone configurations on graphene, such as planar, V-shaped, and Λ-shaped configurations, exhibit distinct energy profiles between ozone and graphene[14]. The presence of extra carriers may modify the ozone configuration and its energy, thereby affecting the ozone affinity for graphene.

Additionally, the polarity of the stray electrostatic field from the gate voltage may contribute to the asymmetric adsorption rate. However, since the graphene layer is grounded, the electric field above the graphene layer is minimal. Nevertheless, the precise mechanisms underlying these phenomena remain incompletely understood and require further investigation.

As a summary, ozone adsorption at cryogenic temperature is investigated. In combination of ultimate sensitivity of graphene and adsorption at cryogenic temperature, a significant modulation to graphene is observed as ozone are absorbed on graphene. The dominating adsorption can be ascribed to physisorption. The significant hole doping into graphene indicates strong Van der Waals interaction between graphene and ozone molecules, requiring future investigation. Furthermore, ozone adsorption has been observed to be tunable by gate voltage, although the mechanism requires further study. The gate tunable asymmetric adsorption may play an important role in the understanding to electrochemical reaction, such as electrocatalysis. The finding of the gate dependence of adsorption will shed new light on making device at atomic scale[9,34,35] and material synthesis[35], such as molecular beam epitaxy (MBE). Typically, the molecules adsorption in MBE is controlled by substate temperature, and the gate voltage may offer an additional controllable parameter, playing an important role in those epitaxial growth with narrow temperature windows and complex growth with multiple elements, such as complex oxide materials. The findings indicate that physically adsorbed molecules enable reversely functionalizing materials, thereby adding a novel avenue to modulate quantum materials and devices.

## Methods:

The graphene, h-BN flakes were exfoliated onto 285-nm SiO2/Si substrates by using the Scotch tape method. Before exfoliation, all SiO2/Si substrates were pre-cleaned by plasma. The layer thickness is determined by Raman spectroscopy and Atomic Force Microscopy. The heterostructure devices were fabricated using the pick-up technique[29]. The Cr/Au electrodes were fabricated using conventional electron-beam lithography combined with thermal evaporation deposition. Four probe transport measurements were carried out in a homemade cryogenic stage.

Figure Caption

**Figure 1|Principle of the ozone adsorption on graphene. a,** Principle of the common adsorption process. **b,** A cartoon of the physical and chemical adsorption of ozone molecules on graphene. **c,** Topview of the graphene/h-BN/graphite device. **d,** Optical image of the graphene flake prepared using a dry transfer technique. **e,** Raman spectra of graphene.

**Figure 2|Ozone adsorption time and temperature dependence. a, c,** Resistance measurement of ozone and oxygen treatment of different time. **b,** Temperature dependence of ozone exposure. Black line and red line denotes exponential and linear fitting of carrier density respectively, blue dots are experimental data from J.Phys. Chem. C 2023, 127, 22015−22022. **d,** Schematic of the ozone physical adsorption process. **e, f,** Comparison of the resistance change of electron-doped and hole -doped graphene after ozone adsorption.

**Figure 3|Gate voltage dependence of the ozone adsorption. a,** The R-$V_G$ curves of graphene after ozone exposure at different gate voltages. **b,** The change of $V_{CNP}$ and the $\Delta V_{CNP}$ after different sequences, blue circles denote the $V_{CNP}$ and orange line denote $\Delta V_{CNP}$. **c,** Gate dependence of the $\Delta V_{CNP}$ and $\Delta n$.

**Figure 4|The asymmetric effect induced by opposite gate voltages and the reducibility of ozone adsorption. a-d,** Comparison of the $V_{CNP}$ shift after ozone exposure at different gate voltage as graphene atop on SiO2. The $V_{CNP}$ shifts in **a** is larger than in **c**. Plot in **b** and **d** show schematic adsorption processes under opposite gate voltages. **e, f,** The evolution of $V_{CNP}$ and $\Delta V_{CNP}$ of two devices in different sequences. Device 1 is treated twice as a comparison shown in **f**.

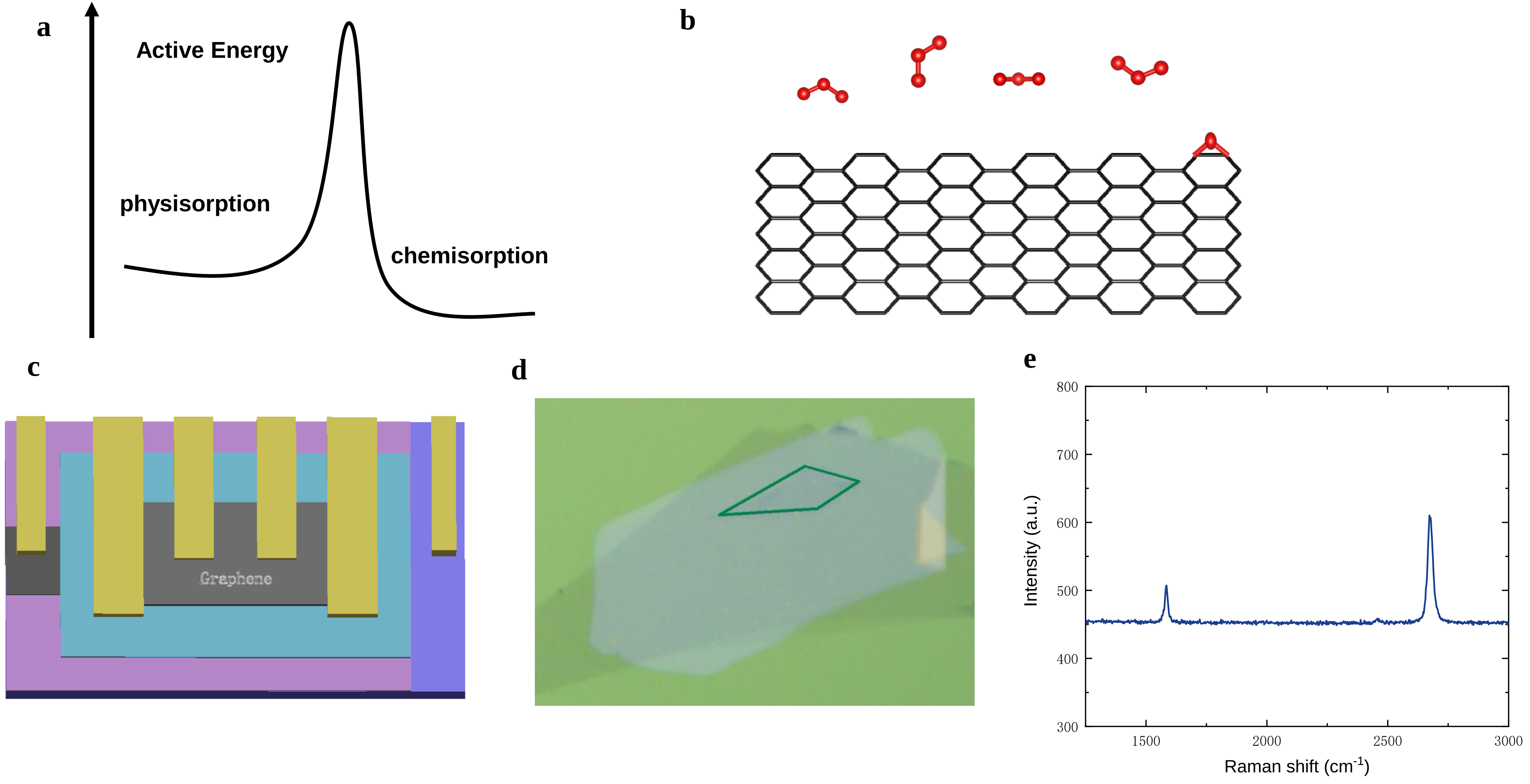

**Figure 1**

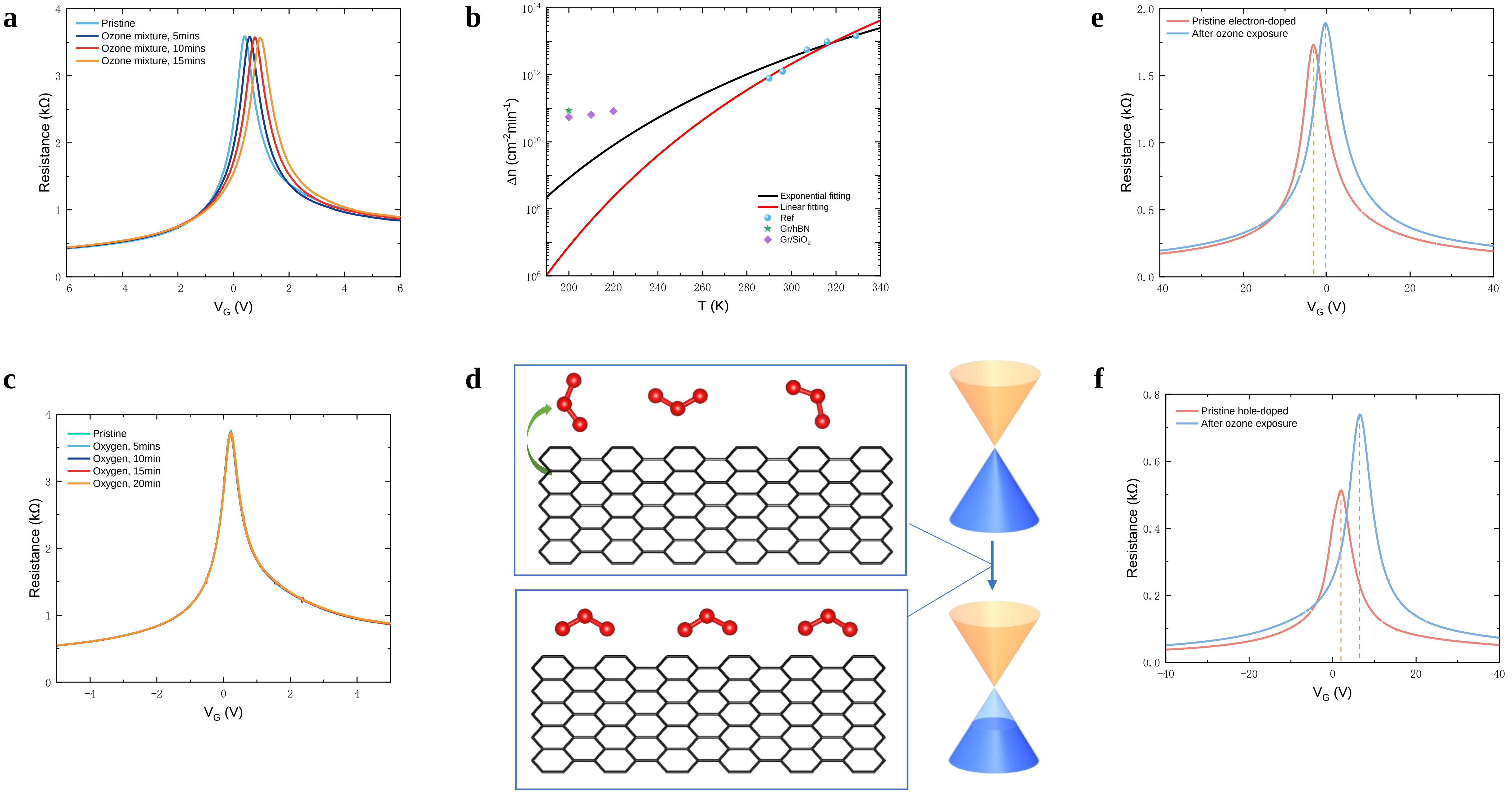

**Figure 2**

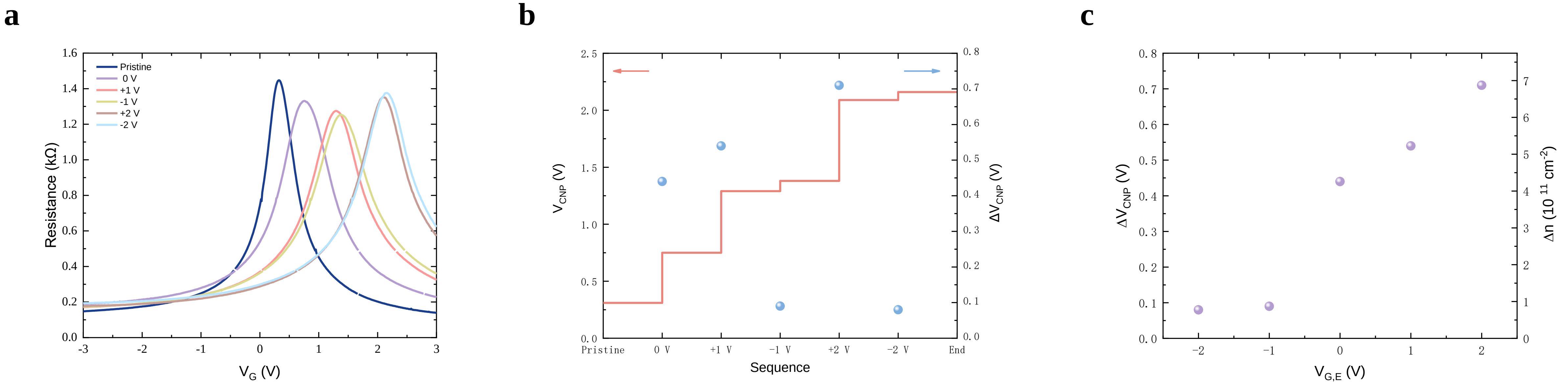

**Figure 3**

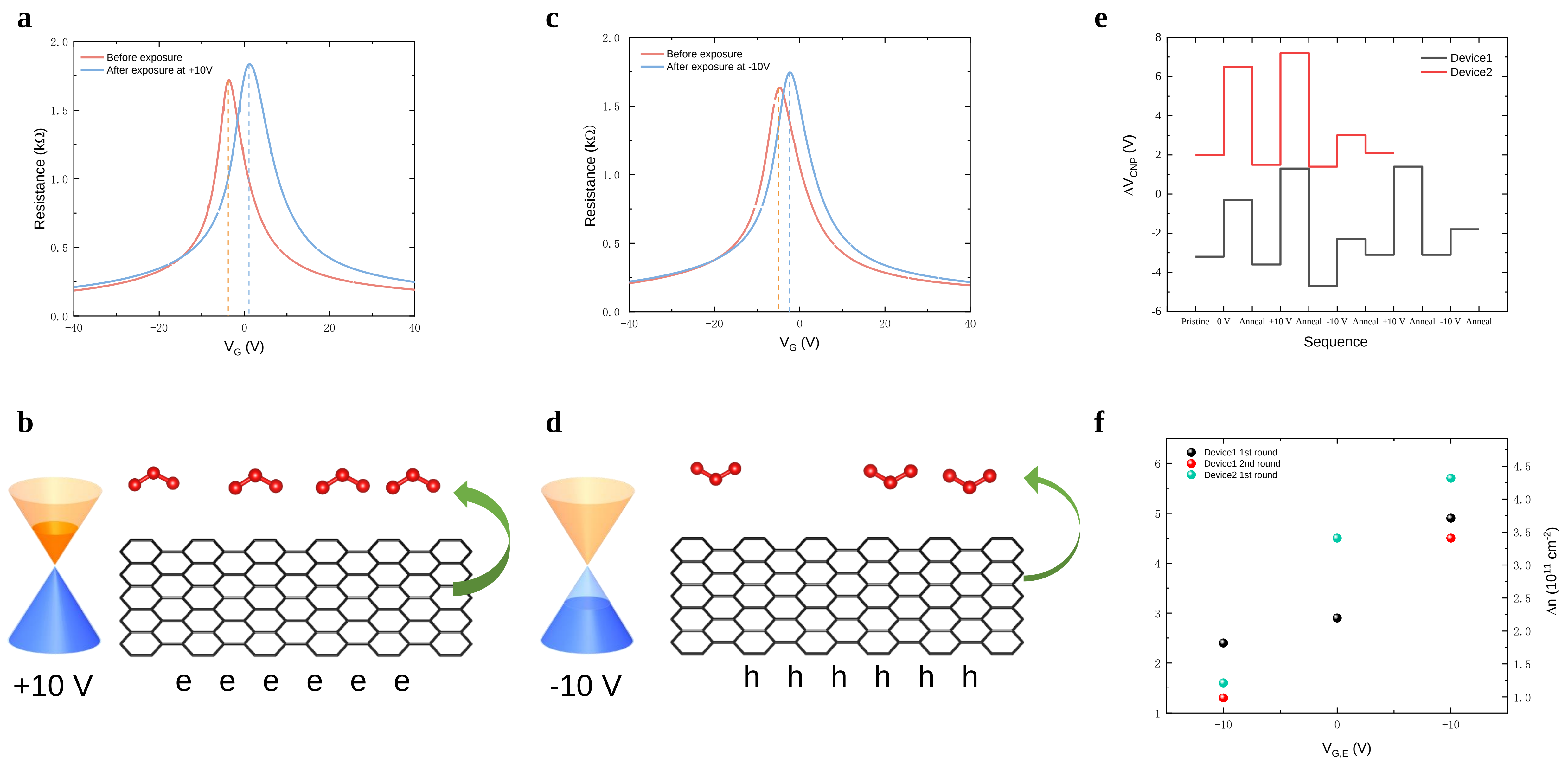
a
Before exposure
After exposure at +10V
Resistance (kΩ)
$V_G$ (V)
b
+10 V
e e e e e e
c
Before exposure
After exposure at -10V
Resistance (kΩ)
$V_G$ (V)
d
-10 V
h h h h h h
e
Device1
Device2
$\Delta V_{CNP}$ (V)
Pristine 0 V Anneal +10 V Anneal -10 V Anneal +10 V Anneal -10 V Anneal
Sequence
f
Device1 1st round
Device1 2nd round
Device2 1st round
$\Delta n$ ($10^{11}$ $cm^{-2}$)
$V_{G,E}$ (V)

**Figure 4**